# $In_2O_3$ doped with hydrogen: electronic structure and optical properties from the pseudopotential Self-Interaction Corrected Density Functional Theory and the Random Phase Approximation

Juan J. Meléndez[1,2] and Małgorzata Wierzbowska[3]

[1] Department of Physics, University of Extremadura.

[2] Institute for Advanced Scientific Computing of Extremadura (ICCAEX)

Avda. de Elvas, s/n, 06006, Badajoz (Spain)

[3] Institute of Physics, Polish Academy of Sciences.

Al. Lotników 32/46, PL-02-668, Warsaw (Poland)

**Abstract**

We discuss the applicability of the pseudopotential-like self-interaction correction (pSIC) to the study of defect energetics and electronic structure of $In_2O_3$. Our results predict that substitutional (at oxygen sites) and interstitial (at antibonding positions) hydrogen, as well as oxygen vacancies with charges +1 and +2, are stable configurations of defects in cubic $In_2O_3$; the former form shallow levels (only as substitutional defects), whereas the latter form deep levels. The band structure calculated with the pSIC shows an excellent agreement with experimental data. In particular, the gap for defect-free $In_2O_3$ is 2.85 eV, which compares fairly with the experimental range 2.3-2.9 eV. The pSIC results also point to a change from indirect to direct gaps depending on doping. In relation to the optical properties, obtained within the random phase approximation, it is shown that they are mostly affected by the presence of oxygen vacancies.

I. INTRODUCTION

Despite they are known since the 50's [1], the so-called transparent conducting oxides (TCOs) have become increasingly more relevant during the last decade because of their interesting technological applications. A short review of them may be found in [2] and references therein; for their interest, we will just mention that potential uses in electronics have been pointed out when thin layers of TCOs are put into contact [3,4]. Amongst TCOs, indium oxide ($In_2O_3$) outstands because it has a good combination of conductivity and optical transparency, which, in addition, is not affected much by doping [5]. Most of the interest of this system has been focused on tin-doped $In_2O_3$ (indium tin oxide, ITO), for which a number of papers about structure, conduction and optoelectronic properties is available [6]. Recently, doping with hydrogen has also received interest because it yields $n$-type behavior with enhanced electron mobility ($n \approx 2 \cdot 10^{20}$ cm$^{-3}$ and $\mu > 120$ cm$^2 \cdot$V$^{-1} \cdot$s$^{-1}$ [7], compared with the typical range $n = 10^{19} - 10^{21}$ cm$^{-3}$ and $\mu = 20 - 40$ cm$^2 \cdot$V$^{-1} \cdot$s$^{-1}$ reported in thin films of ITO [8]) keeping a good balance between transparency and conductivity, at least within some range of wavelengths [9]. In the end, hydrogen is very difficult to remove from any crystal growth environment and, in addition, forms a strong bond with oxygen, which may act as a driving force for its incorporation to the lattice of $In_2O_3$ during crystal growth. Evidence has also been found that co-doping can even improve the conductivity of hydrogen-doped $In_2O_3$ [10].

Surprisingly, despite its potential relevance as donor dopant, the studies about the role of hydrogen on the optoelectronic performance of $In_2O_3$ are not too numerous. From the point of view of numerical simulation, Limpijumnong and co-workers used the density functional theory (DFT) in a pioneer paper about energetics of defects in hydrogen-doped $In_2O_3$ for some configurations of defects: interstitial hydrogen with charges 0, +1 and -1, substitutional hydrogen at oxygen sites and oxygen vacancies with charges 0 (neutral), +1 and +2 [11]. They found that only $H_i^{\cdot}$ at antibonding sites, $H_O^{\cdot}$ at oxygen sites and $v_O$ in the +2 charge state are stable for all the values of the Fermi energy within the bandgap. They also concluded that hydrogen is a shallow donor in hydrogen-doped $In_2O_3$, regardless its particular location within the lattice, whereas oxygen vacancies are deep donors. However, these conclusions are stated from energetic considerations, since the authors do not report the band structure of hydrogen-doped $In_2O_3$. In addition, while their findings agree reasonably well with the experimental evidence [7,12], one does not feel fully satisfied with the use of the DFT+U method, in which the particular exchange-correlation functional is modified by a Hubbard-like term modulated by a semiempirical parameter $U$. The DFT+U method is currently used (successfully in many cases) in strongly correlated systems (such as magnetic materials or, in general, solids comprising atoms with semifilled $d$ shells), but the comparability of the results with experimental data is quite sensitive to the value of $U$, which is not always supported by a sound physical basis. In general, one would like to avoid DFT+U in a fully *ab initio* study.

Part of the incapacity of DFT to describe correlated systems lies in that both the LDA and the GGA functionals contain a spurious self-interaction effect (i.e., the interaction of an electron with the full potential created by itself) which is particularly severe in systems with spatially localized electrons. Strategies have been proposed to correct the self-interaction effect from the formalism of DFT (see, for instance, [13] and references therein). Among these, we have chosen the realization of Filippetti and Spaldin [13], which is particularly useful when strongly localized and hybridized electron coexist, as will be our case. Within this framework, the exchange-correlation functional is corrected by a set of orbital occupation numbers calculated self-consistently. As a result, an additional non-local projector accounting for self-interaction appears in the Kohn-Sham equations. If pseudopotentials are used to solve them, they contain then a self-interaction correction by construction, thence the name of pseudopotential-like self-interaction correction (pseudo-SIC, or simply pSIC). The pSIC method by Filippetti and Spaldin successfully corrects some flagrant failures of the "standard" DFT in what respects to the atomic and electronic structures of silicon, wide-band gap insulators, transition-metal-based oxides and manganese oxides [13-16]. We note here that the pSIC method has the obvious advantage to require a computational cost similar to that for a DFT+U calculation, much less than other techniques based on the many-body theory.

This study has then a double objective. The first one is to test the suitability of the pSIC (in the Filippetti and Spaldin realization) to deal with the electronic structure and optical properties of defect-free $In_2O_3$. We will show that pSIC corrects the standard DFT results in what respects to the band structure of this system at a moderate computational cost. The second goal is to extend the study by Limpijumnong and co-workers about the effect of hydrogen doping in $In_2O_3$. In particular, the band structure and optical properties of hydrogen-doped $In_2O_3$ for several configurations of defects will be investigated within the DFT framework corrected with pSIC. The results will be compared with available experimental data and conclusions will be stated.

## II. METHODOLOGY

The defect-free $In_2O_3$ supercell (space group $Ia\bar{3}$, $a = 10.14$ Å) contains 16 formula units with a total of 80 atoms; table I shows the details of the crystal structure. Limpijumnong and co-workers demonstrated that substitutional hydrogen at O-sites as well as interstitial hydrogen at antibonding sites exhibit positive charge, whereas it is either neutral or negatively charged nearby the cations [11]. Therefore, five configurations of hydrogen defects were created from the ideal defect-free unit cell: interstitial hydrogen at $8a$, at $16c$ with $x = 0.5$ and at an antibonding position [*cf.* fig. 1], and substitutional hydrogen at an oxygen site. Thus, we considered the $H_O^{\cdot}$, $H_i$-AB, $H_i'$-$16c$, $H_i^{\times}$-$16c$, $H_i'$-$8a$ and $H_i^{\times}$-$8a$ configurations. In hydrogen-doped $In_2O_3$, it has been pointed out that oxygen vacancies could appear because of the dissociation of $H_i$ into $H_O$ [11]. Therefore, we considered oxygen vacancies with different charge states, namely $v_O^{\cdot\cdot}$, $v_O^{\cdot}$ and $v_O^{\times}$, as relevant defects as well.

The *ab initio* calculations were carried out by DFT with and without pSIC under the linear density approximation (LDA) for the exchange-correlation functional. The DFT calculations were performed by the Quantum Espresso code [17] modified with the pSIC realization of Filippetti and Spaldin [14]. Norm-conserving scalar-relativistic pseudopotentials were used for all the atoms, with $d$ semicore electrons for indium. After convergence tests, the energy cutoff was set to 130 Ry, and a Monkhorst-Pack grid of $4 \times 4 \times 4$ was used to sample the first Brillouin zone. In order to get a smooth representation of the band structure, a set of calculations was performed on a Monkhorst-Pack grid of $8 \times 8 \times 8$, and the energies were then extrapolated to some high-symmetry directions of the first Brillouin zone.

The coordinates of all the atoms within each supercell were allowed to vary until the total Hellmann-Feynman force decreased below 0.02 eV/Å. The formation energy for each defect configuration of charge $q$ was then calculated as

$$E_f(d^q) = E_{tot}(d^q) - E_{tot}(\text{bulk}) + \sum_i n_i \mu_i + q\varepsilon_F, \qquad (1)$$

where $E_{tot}(d^q)$ and $E_{tot}(\text{bulk})$ hold for the total energy of the defective and ideal supercells, respectively, $n_i$ is the number of atoms of the $i$-th species which must be added ($n_i < 0$) or removed ($n_i > 0$) to create the actual defect configuration, $\mu_i$ is the chemical potential of the $i$-th species and $\varepsilon_F$ is the Fermi energy. The chemical potentials for each species depend on their particular chemical environment. Here, we took $\mu_H$ and $\mu_O$ equal to half the total DFT energy of isolated hydrogen and oxygen molecules, respectively.

The dielectric function in the optical limit (that is, for zero transferred momentum) of defect-free and defective cells was calculated from the respective Kohn-Sham eigenvalues under the random-phase approximation (RPA, [18]) and at time-dependent DFT level using the adiabatic LDA functional (ALDA, [19]). Both approaches may be affected by the so-called local field effects, which account for the off-diagonal elements of the inverse dielectric function. In our calculations, these local field effects did not have any influence and, therefore, all the results shown here were taken neglecting these effects. In addition, the RPA and ALDA yielded virtually the same results in all cases; we are showing only those for RPA for simplicity. As will be shown below, the supercell contaning an oxygen vacancy with charge +1 exhibits metallic character. In this case, the contribution of intraband transitions to the dielectric function can be relevant. In the limit of zero transferred momentum, these transitions may be described by a Drude-like term depending on the plasma frequency of the system [20]. The plasma frequency was taken as real and equal to 85 meV [21], which was the estimated value for defect-free In$_2$O$_3$. From the dielectric function, the absorption coefficient was calculated as function of the energy as

$$\alpha(\varepsilon) = \frac{2\kappa(\varepsilon)\varepsilon}{\hbar c}, \qquad (1)$$

where

$$\kappa(\varepsilon) = \frac{1}{\sqrt{2}}\left[-\varepsilon_1(\varepsilon) + (\varepsilon_1^2(\varepsilon) + \varepsilon_2^2(\varepsilon))^{1/2}\right]^{1/2}, \qquad (2)$$

and $\varepsilon_1$ and $\varepsilon_2$ hold for the real and imaginary parts of the dielectric function, respectively. These calculations were performed by the YAMBO code [22].

### III. DEFECT FORMATION ENERGIES

Fig. 2 shows the calculated formation energies for the configurations of defects as functions of the Fermi energy. The vertical straight line to the right denotes the calculated gap for defect-free In$_2$O$_3$, with the Fermi energy taken as zero at the valence band maximum (VBM). The figure indicates that $H_i^\times$ and $H_i'$ are unstable in any case, $H_i^{\cdot}$ at antibonding sites being the most stable defect for all the values of the Fermi energy. $H_O^{\cdot}$ is a stable defect too. Its formation energy equals zero for $\varepsilon_F = 2.70$ eV, but the difference with the calculated gap (2.85 eV) is too small to think that it could be unstable. Anyway, the difference between the formation energies of $H_i$-AB and $H_O^{\cdot}$ is around 4 eV for all the values of the Fermi level, which suggests that substitutional hydrogen is unlikely to exist. Thus, the pSIC results support the idea that hydrogen is a donor agent in In$_2$O$_3$, as was already pointed out by Limpijumnong and co-workers [11].

As for oxygen vacancies, only charged ones are stable, with $v_O^{\cdot\cdot}$ exhibiting formation energy similar to that for $H_i^{\cdot}$ within the bandgap. On the contrary, $v_O^{\cdot}$ has a higher formation energy, which equals zero for $\varepsilon_F = 2.46$ eV; this value is still close enough to the calculated gap to consider it as stable. Our results about formation energies agree with those reported by Limpijumnong and co-workers [11]; the quantitative differences arise mainly because we are using self-interaction corrections in contrast to the LDA+U approach used by them.

## IV. CRYSTAL STRUCTURE OF DEFECTIVE SUPERCELLS

$In_2O_3$ is isomorphic with bixbyte [$(Mn,Fe)_2O_3$], with two non-equivalent sites for indium atoms at Wyckoff positions 8*b* and 24*d*, as shown in Figs. 3a and 3b. Indium atoms at 8*b* sites (hereafter referred to as In-8*b*) are six-fold coordinated with oxygen, with all the bonds having the same length and the angles ranging between 80º and 100º, approximately. We may consider then that In-8*b* have a distorted octahedral coordination with its oxygen neighbors. Indium at 24*d* sites (hereafter referred to as In-24*d*) is six-folded coordinated with its oxygen neighbors as well. In this case, the bond lengths are equal only for pairs of oxygen defining a plane with the In-24*d* ion. The angles also range roughly between 80º and 110º, so that the coordination is trigonal prismatic. Oxygen atoms locate at Wyckoff positions 48*e* in the bixbyte structure. Each of them is surrounded by three In-24*d* and one In-8*b* in a distorted tetrahedral coordination in which all the bond lengths are dissimilar. The angles vary between around 100º and 125º in this case.

Figs. 4a to 4d show the bond lengths and angles for the stable configurations of defects. In most cases, the inclusion of the defect yields bond lengthening and angles wrinkling. The $H_i^-$-AB configuration yields the most severe distortion, with bonds lengthening more than 10% with respect to the ideal arrangement and angles wrinkling up to the 7.7%. This fact agrees with that hydrogen atoms occupying antibonding positions tend to enlarge the bond distance with the corresponding host atom, as has been identified in ZnO [23] and GaN [24]. Note, however, that the antibonding positions are not the most stable ones for interstitial hydrogen in the latter system, contrarily to what happens in $In_2O_3$. We could argue that the reason for this is the tendency of hydrogen to form H-O bonds. Actually, severe lattice distortions for interstitial hydrogen at antibonding positions have also been reported in ZnO [23]. Oxygen vacancies also push outwards the surrounding indium atoms, the distances In-vacancy and the angles being higher for the +2 charge state. This fact was expectable, since the oxygen vacancies act as positive charges. We remark here that, in all cases, the atomic arrangement remained practically unchanged outside the second coordination sphere from each defect, which indicates that the interaction with images of the defects generated by the periodic boundary conditions is negligible. All these findings are in good agreement with those by Limpijumnong and co-workers [11]. Incidentally, we could mention that the atomic positions calculated without pSIC (not shown here for clarity) are very similar to those with pSIC, which indicates that this correction does not play a major role in the structure relaxation.

## V. ELECTRONIC STRUCTURE

### A. Band structure around the gap for defect-free $In_2O_3$

Part of the pSIC band structure of defect-free $In_2O_3$ [namely the five highest valence bands (VBs) and five lowest conduction bands (CBs)] along the line $\Gamma - N\langle 110\rangle - H\langle 100\rangle - \Gamma - P\langle 111\rangle$ is shown in Fig. 5; all the energies are relative to the VBM. The thin dashed line corresponds to the bands obtained from DFT without pSIC. The CB is highly dispersive, whilst the top of the VB is practically flat, with an energy width of 90 meV. The calculated gap is indirect, with the conduction band minimum (CBM) located at the $\Gamma$ point.

One first notices that the fundamental energy gap calculated with the pSIC is 2.85 eV, well above that calculated by uncorrected DFT (1.36 eV) and in very good agreement with the experimental values ranging between 2.3 eV and 2.9 eV [25-33]. The underestimation of the bandgap is inherent to standard DFT, regardless the particular exchange-correlation functional (LDA, GGA) used. In $In_2O_3$, it has been argued that the presence of In $4d$ electrons strengthens the gap underestimation [34]. Indeed, the topmost valence band in this system is formed by hybrid In $4d$ - O $2p$ states (see below). Wei and Zunger found an important *pd* repulsion in II-VI semiconductors with shallow *d* levels within the valence band [35], one of whose effects is to reduce the band gap; the same effect was identified in InN [36]. The most obvious solution to this gap shrinkage is to open rigidly the gap using the so-called scissor operator; however, this procedure just shifts upwards the entire CB to match the correct gap, without changing the structure of the energy levels. The DFT+U method has been used in $In_2O_3$ to overcome this problem too, with results in good agreement with experiments [37]. Contrarily, the pSIC that we have used is explicitly taken into account at every step of the calculation and, therefore, it does not limit itself to a mere shift of the band structure.

Another interesting issue about the band structure of $In_2O_3$ is the character of its fundamental gap, which has long been controversial. The sharp absorption edge exhibited by $In_2O_3$ at 3.5–3.75 eV was primarily attributed to a direct gap, although a much weaker absorption edge within the range 2.3–2.9 eV suggested that the gap could actually be indirect [27,38]. Recent measurements by X-ray photoelectron spectroscopy (XPS) could be interpreted in terms of a fundamental direct gap in the range 2.7-2.9 eV [31,39], as confirmed by angular-resolved photoelectron spectroscopy (ARPES) [32,33], but also in terms of an indirect gap [30].

*Ab initio* calculations are apparently controversial at this respect as well. Mryasov and Freeman used the linear muffin-tin orbital method with atomic sphere approximation and found a direct gap of about 1 eV in defect-free $In_2O_3$ [40]. Medvedeva used a full-potential augmented wave method to reach a direct gap as well, of about 1.16 eV [41]. Karazhanov and co-workers also underestimate the gap, which identify as direct [42]. Contrarily, Erhart and co-workers [37], Walsh and co-workers [26] and Fuchs and co-workers [34] find and indirect gap for $In_2O_3$ with DFT+U and many-body perturbation theory, respectively. This is also our case. This apparent discrepancy between simulations and experiments may be matched up by noticing that when the gap is identified as indirect, the width of the topmost valence band is small (about 50 meV in [37] and [26], less than 10 meV in [34] and 90 meV in this work), much lower than the differences between the aforementioned absorption edges. An extremely flat upper valence band is consistent with the ambiguous character of the fundamental optical gap, and the weak absorption edge measured at 2.3-2.9 eV may be considered as a direct forbidden / indirect allowed dipolar transition [26]; note that a direct forbidden transition is consistent with the *ab initio* calculations. In other words, computer simulation does not justify the sharp indirect gap originally thought to exist in $In_2O_3$. In addition, the level of doping strongly affects the analysis of the experimental data, as will be discussed below. The main effect is the Burnstein-Moss shift appearing in doped semiconductors where the Fermi level locates within the conduction band. It is interesting to realize that the fundamental gap has been identified as indirect in high-purity samples [30]. Finally, apart from dopant impurities, there are other defects (like dislocations) which may well affect the experimental absorption spectra but which are not explicitly included in computer simulations.

### B. Band structure near the gap of defective $In_2O_3$

Figs. 6a to 6d show the band structure for the four stable configurations of defects; the dashed lines in these figures correspond to the band structure of defect-free $In_2O_3$, calculated within the pSIC scheme. The incorporation of hydrogen produces changes in the band structure near the gap. As for defect-free $In_2O_3$, the CBs are dispersive and the VBs are flat, but with widths exceeding 100 meV in all cases. For the $H_i^{\cdot}$-$AB$ configuration, the incorporation of hydrogen simply yields a roughly rigid shift upward of the whole CB, which increases the gap to 3.01 eV. For $H_O^{\cdot}$, on the contrary, a defect level appears, at 3.46 eV above the VBM. This level is 1.15 eV wide, and reaches its maximum and minimum at the $H$ and $\Gamma$ points, respectively, and its presence yields a significant increment of the bandgap to 4.61 eV. It is interesting to realize that pSIC predicts this defect level to be very shallow; in Fig. 6b, it actually appears slightly overlapping the bottommost CB due to numerical accuracy. In addition, it exhibits essentially the same topography as the bottommost CB in defect-free samples, with extremes at the same points of the first Brillouin zone. Oxygen vacancies also create defect levels within the gap. For both states of charge, the bandgap increases to 4.56 eV. However, both the position and the width of the defect level depend on the charge of the oxygen vacancy. For $v_O^{\cdot}$, the level is located 0.91 eV below the CBM, and it is 1.27 eV wide; for $v_O^{\cdot\cdot}$, the level locates 0.55 eV below the CBM, and its width is 1.50 eV. In any case, the defect levels caused by oxygen vacancies can be regarded as deep. The defect levels are plotted in red in Figs. 6b to 6d.

These results are in relative good agreement with the *ab initio* calculations by Limpijumnong and co-workers [11] and with experimental data by King and co-workers [43]: the main difference compared to our results is that we do not find interstitial oxygen to give a defect level, at least at antibonding positions. Note that these results confirm that the presence of hydrogen as unintentional dopant may explain the high *n*-conductivity of $In_2O_3$, even when it is expected to be defect-free. The high values of conductivity cannot be ascribed to oxygen vacancies, since they form deep (thus affecting optical properties, as will be shown below) levels, as our results demonstrate. Incidentally, we mention that the same conductive transcendence of hydrogen has been identified in many other oxides [44-46].

A second issue about the band structure of defective $In_2O_3$ is that, when defects others than $H_i$-AB are present, the gap becomes direct, as shown in Figs. 6b to 6d. We have pointed out before that the level of doping and the dopant agent could lie behind the controversy about the direct-indirect character of the fundamental gap in $In_2O_3$. The presence of hydrogen could well lie behind this controversy, which exists even in high-purity samples. Anyway, many other studies, with many other dopant agents, should be performed to be conclusive at this respect, of course.

## VI. DENSITIES OF STATES

Fig. 7 plots the total density of states (DOS) for defect-free $In_2O_3$ together with the partial densities (pDOS) obtained projecting the total DOS onto orbital atomic states. As before, the energies are relative to the VBM and the dashed line corresponds to the DOS calculated from DFT without pSIC. The VB has two parts, corresponding to the energy ranges (-18.0) – (-15.4) eV and (-6.3) eV – 0.0 eV. Contribution from indium orbitals (either *s*, *p* or *d*) overlaps with that from oxygen ones in both parts, which indicates that the bonding in $In_2O_3$ has an important covalent component. The bottommost VB is essentially contributed by oxygen $2s$ and In $4d$ electrons (not shown individually here for clarity), with negligible contribution from oxygen $2p$ ones. The main contribution to the topmost VB, on the contrary, comes from oxygen $2p$ and indium $5p + 4d$ electrons. The CB, for its part, arises from the hybridization of In($s + p + d$) electrons with oxygen $2p$ and $2s$ ones, being the first the most important contribution.

Blyth and co-workers measured the width of the topmost VB, essentially triangular shaped, to be $\sim 6$ eV, with an important contribution of the In $4d$ states at its bottom edge [47]; Walsh and co-workers report a topmost VB slightly less than 6 eV wide as well [26]. As for the rest of the valence band, it has been suggested that it consists actually on three regions [42]; however, the pSIC band structure that we are reporting here, with just two regions in the VB, is in much closer agreement with data from soft X-ray emission spectroscopy and absorption and from X-ray photoelectron spectroscopy [48,49], as demonstrated in Fig. 8.

Figs. 9a to 9d plot the DOS and pDOS for the stable configurations of defects. The features of the DOS are different depending on whether hydrogen was incorporated to the crystalline lattice or not. For $H_i$- $AB$ two narrow defect levels appear, one located at around -20.0 eV, below the bottommost VB, and the second at around -7.4 eV, below the topmost one. These defect levels arise from the hybridization of indium with oxygen $2s$ and oxygen $2p$ electrons, respectively. For substitutional hydrogen $H_O^{\cdot}$ another defect level is apparent, at -6.8 eV, arising by the hybridization of indium and oxygen $2p$ electrons. The shallow defect level within the gap does (which does not appear unresolved due to numerical inaccuracies of our calculations, in our opinion) arises by the hybridization of indium levels and oxygen $2p$ levels, with a small contribution from oxygen $2s$.

In figs. 9c and 9d, the deep defect levels for the $v_O^{\cdot}$ and $v_O^{\cdot\cdot}$ configurations appear clearly resolved within the gap. In both cases, the main contribution to these levels comes from In electrons at low energies, but the hybridization with O $2s$ and $2p$ electrons becomes important at higher energies. For $v_O^{\cdot}$ vacancies, the Fermi level is 3.45 eV, which lies within the impurity level; a strong enhancement in the conductivity would be expected if these defects were likely to appear. For $v_O^{\cdot\cdot}$, on the contrary, the Fermi level lies below the shallow impurity level. Therefore, it is not expected any effect on the conductive behavior (an aspect which is still under debate, see for instance Ref. [2] and references therein), but should affect the optical properties of these samples within the optical region of the spectrum.

## VII. DIELECTRIC FUNCTION AND OPTICAL PROPERTIES

As we have already mentioned, pure $In_2O_3$ is transparent in the optical range of the spectra, with a strong absorption edge in the UV range. Figs. 10 plot the imaginary part of the dielectric function $\varepsilon_2$ (a) and the absorption coefficient $\alpha$ (b) for defect-free $In_2O_3$ as functions of the energy; as before, the dashed line corresponds to the results from DFT without pSIC. Because of the cubic symmetry, the diagonal components of the dielectric function are equal within the numerical accuracy; Fig. 10a shows then the imaginary part of the average of $\varepsilon_{xx}$, $\varepsilon_{yy}$ and $\varepsilon_{zz}$. The $\varepsilon_2(\varepsilon)$ function calculated with pSIC takes on values close to zero within the visible range (shaded rectangles in Figs. 10), as accordingly $\alpha(\varepsilon)$ does, in good agreement with the optical transparency of $In_2O_3$. On the contrary, the results from DFT without pSIC yield non-negligible values in the optical range. This fact is primarily attributable to the wrong gap predicted by standard DFT. Besides, since not only the gap but the entire conduction band is affected by the self-interaction correction, the structures of the $\varepsilon_2(\varepsilon)$ and $\alpha(\varepsilon)$ curves also change with respect to those calculated by DFT without pSIC. In particular, in the latter case one observes a set of peaks, particularly sharp in the range 2.4-2.7 eV, which are lack in pSIC calculations.

Fig. 10b includes a comparison of our results for the absorption coefficient of defect-free $In_2O_3$ with experimental data reported in the literature [27,30,50]. The pSIC results are in quite good agreement with those by Hamberg and co-workers up to around 4.0 eV [50], but are up to one order of magnitude higher than those by Weiher and Ley [27] and Irmscher and co-workers [30]. We do not have any satisfactory explanation about this discrepancy between the different sets of data plot in Fig. 10b. In principle, one could think about a temperature effect; however, the experimental data were all collected at room temperature, and still differ to each other. Differences in the layer thicknesses are unlikely to affect the results either, since the absorption coefficient is not too sensitive to the thickness [30]. As before, different degree of impurity may also be affecting the results.

Fig. 11 displays the absorption coefficient as function of the energy for the stable configurations of defects. This plot indicates that hydrogen does not compromise the transparency of $In_2O_3$ regardless its character (interstitial or substitutional). On the contrary, the absorption coefficient is significantly higher within the visible region when oxygen vacancies are present. These defects affect a region of the spectrum that depends on their particular charges. For the $v_O^{\cdot}$ configuration, the absorption is higher at low energies; the absorption curve has in this case a peak structure (a peak at 1.94 eV is clearly observed) that may be attributed to intraband transitions and to the impurity level within the gap mentioned above. For the $v_O^{\cdot\cdot}$ configuration, the absorption increases with the energy.

## VIII. CONCLUSIONS

The above paragraphs demonstrate that the pseudopotential-like self-interaction correction, as formulated by Filippetti and Spaldin [13], can be successfully used to study the electronic structure and optical properties of $In_2O_3$-based materials. In the defect-free system, the pSIC yields significantly better values for the band gap than standard DFT, which compare fairly well with the experimental data available. As a result of our calculations, we point out that the controversial direct – indirect character of the gap in $In_2O_3$ could be related to the level of doping, since the gap is always direct when defects (at least, those stable defects studies herein) are present. The calculated DOS and the optical absorption spectra within the RPA are comparable with experiments as well.

In defected $In_2O_3$, the pSIC predicts that donor hydrogen (either interstitial at antibonding sites or substitutional at oxygen ones) as well as oxygen vacancies with charges +1 and +2 are stable defects. Substitutional hydrogen forms shallow levels that may well affect the electrical conductivity, whereas oxygen vacancies form deep levels affecting mostly optical properties.

Finally, we remark that the pSIC demands relatively few computational resources compared to more sophisticated approaches, yielding essentially the same results in what respects to the electronic structure. The combination of good accuracy and computational cheapness is important in the description of other TCOs, such as Ga- or Al-doped ZnO (GZO and AZO, respectively), which are alternatives to the relatively expensive devices based on $In_2O_3$ and are currently used in industry [51].

ACKNOWLEDGMENTS

This work was founded by the Spanish Ministry of Economy and Innovation under Grant no. MAT2012-38205-C02-02. Computational support by Dr. Carlos J. García-Orellana, from the University of Extremadura and the Institute for Advanced Scientific Computing of Extremadura has been of great help. MW acknowledges Project. No. 2013/11/B/ST3/04041 by the National Science Centre in Poland.

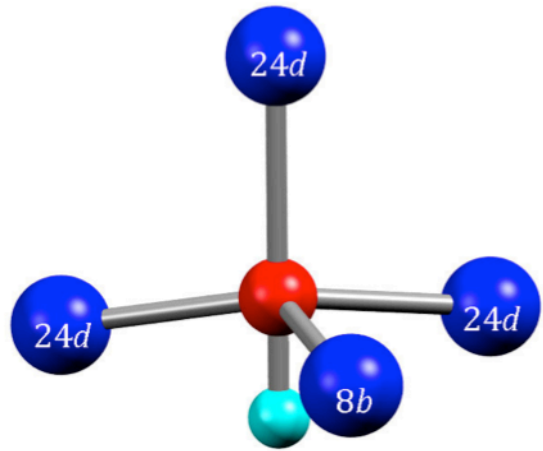

FIG. 1: Location of the hydrogen atom in the $H_i$-AB configuration. In these and subsequent figures, indium and oxygen atoms are represented by blue and red spheres, respectively. The corresponding Wyckoff position of indium atoms is given inside each sphere. Hydrogen appears in light blue.

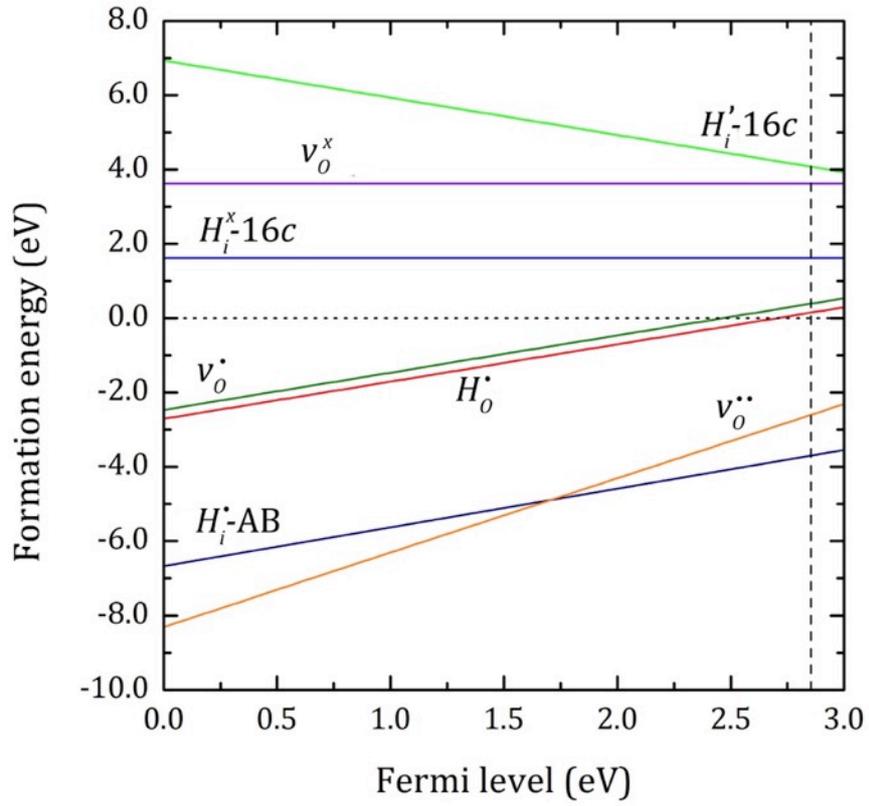

FIG. 2: Formation energies for the configurations of defects as functions of the Fermi level. The dashed line to the right denotes the gap of defect-free $In_2O_3$.

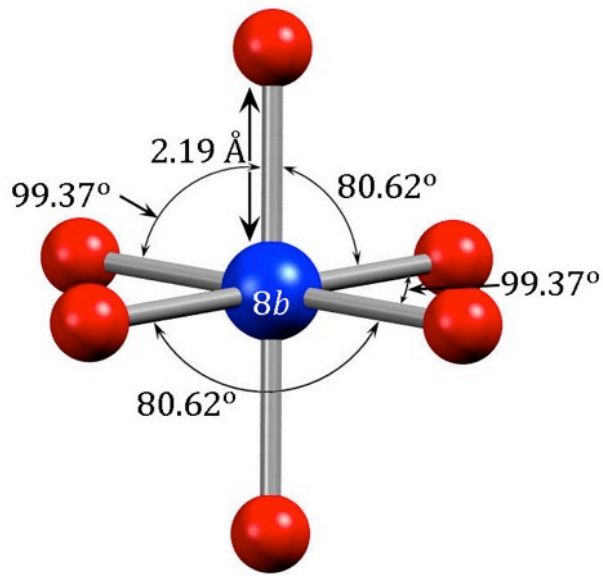

(a)

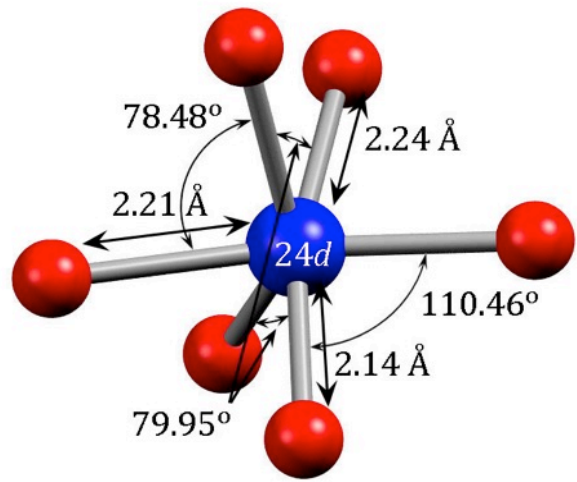

(b)

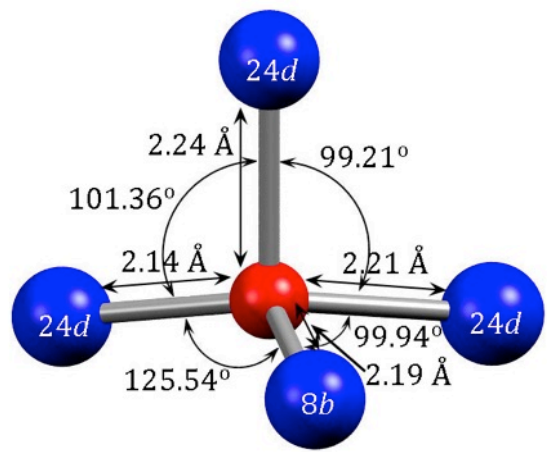

(c)

FIG. 3: Atomic environments of indium atoms, at 8*b* (a) and 24*d* (b) Wyckoff positions, and of oxygen (c).

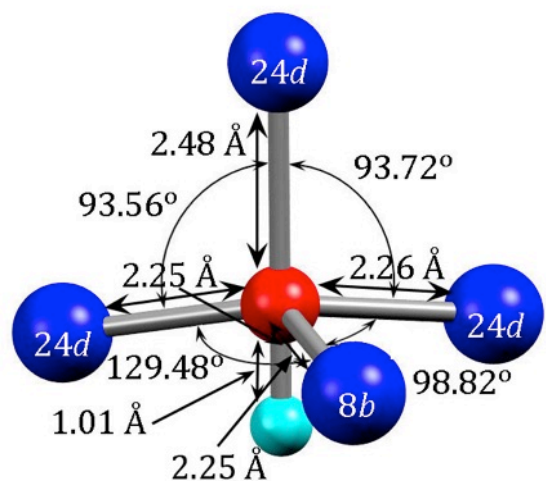

(a)

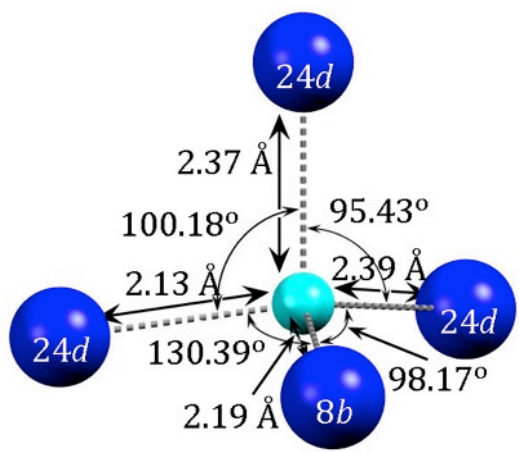

(b)

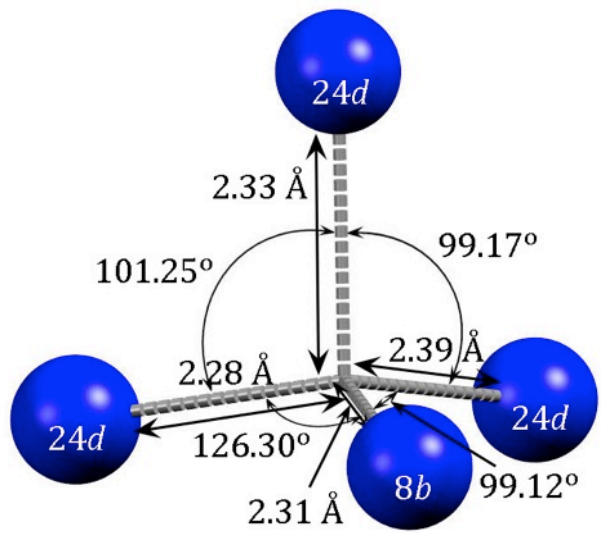

(c)

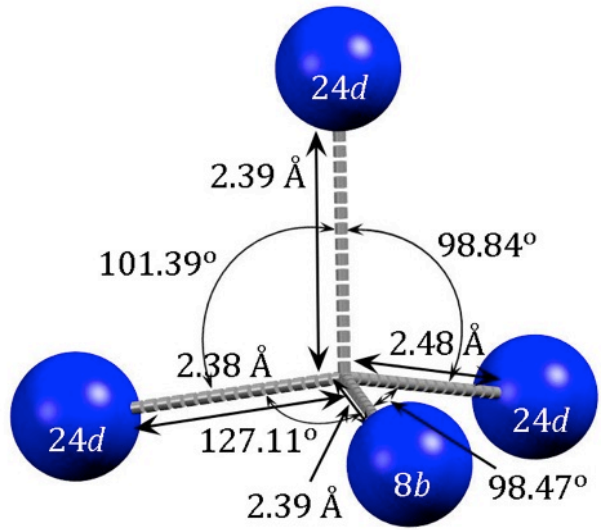

(d)

FIG. 4: Atomic environments for the $H_i^{\cdot}$-AB (a), $H_O^{\cdot}$ (b), $v_O^{\cdot}$ (c) and $v_O^{\cdot\cdot}$ (d) defect configurations after relaxation.

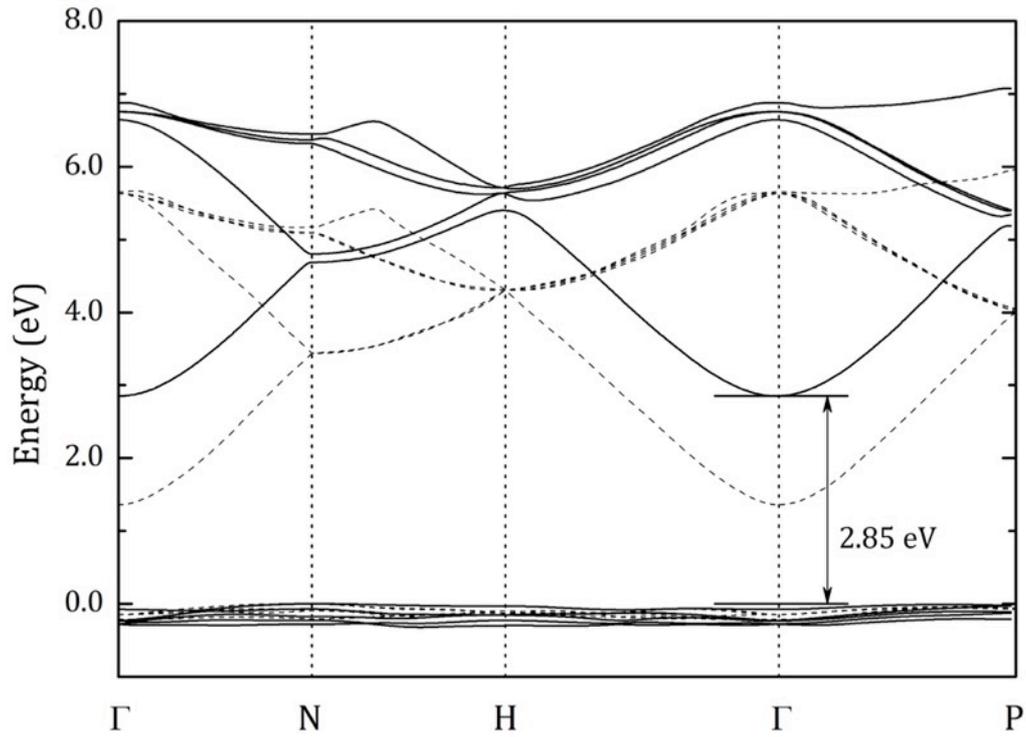

FIG. 5: Band structure of defect-free $In_2O_3$ along some high-symmetry directions of the first Brillouin zone calculated with (solid lines) and without (dashed lines) self-interaction corrections. The arrows denote the bandgap.

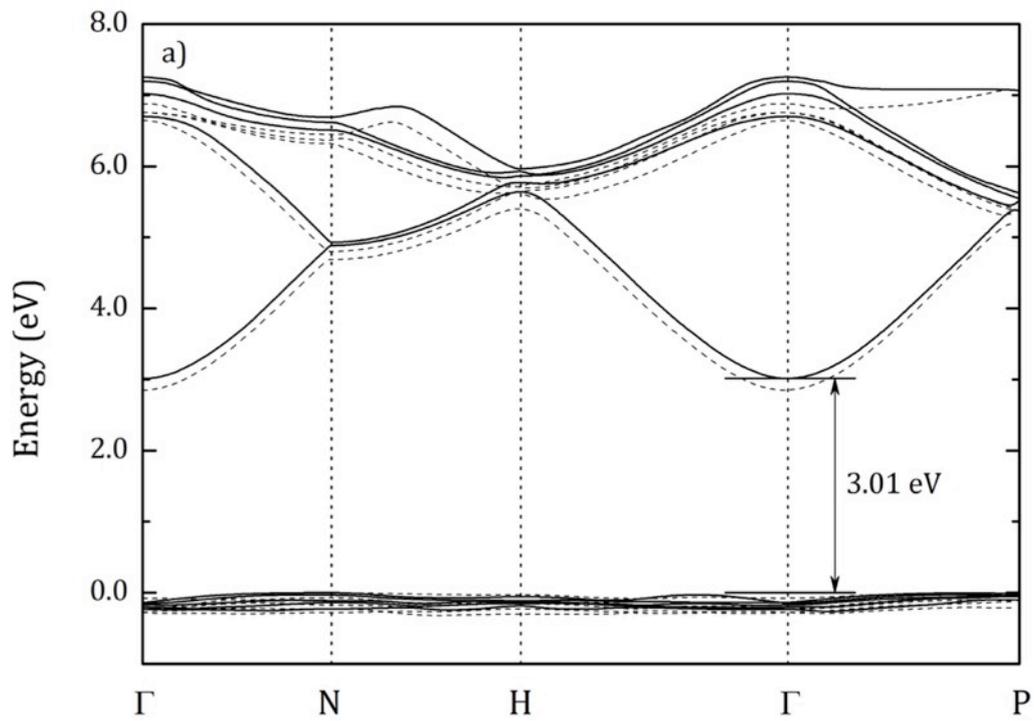

(a)

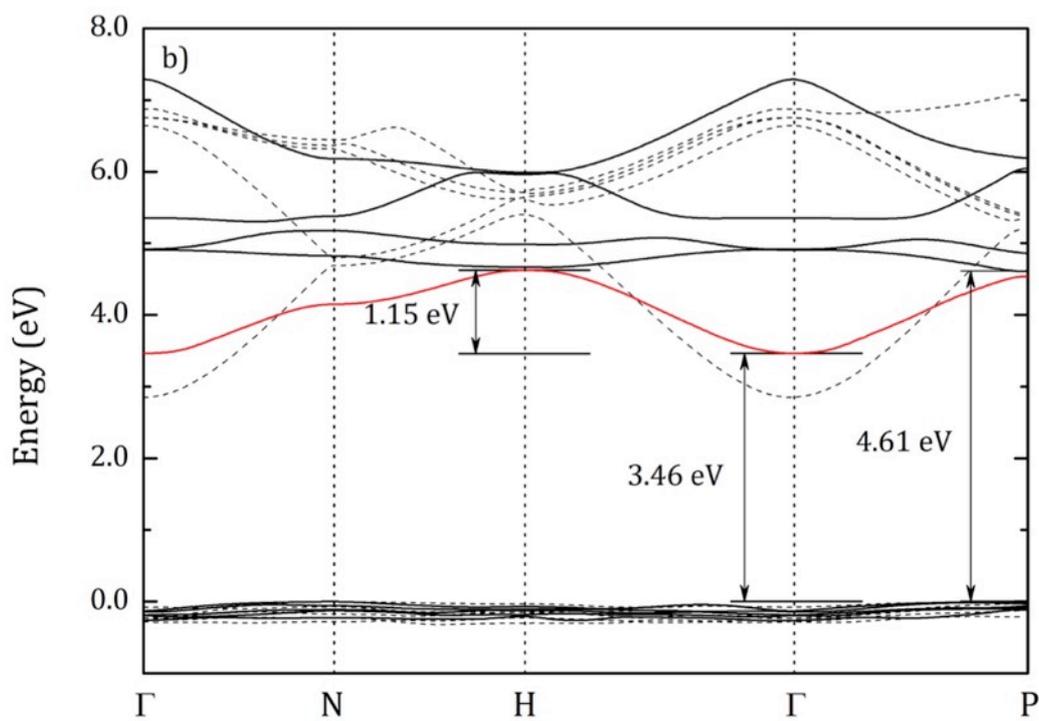

(b)

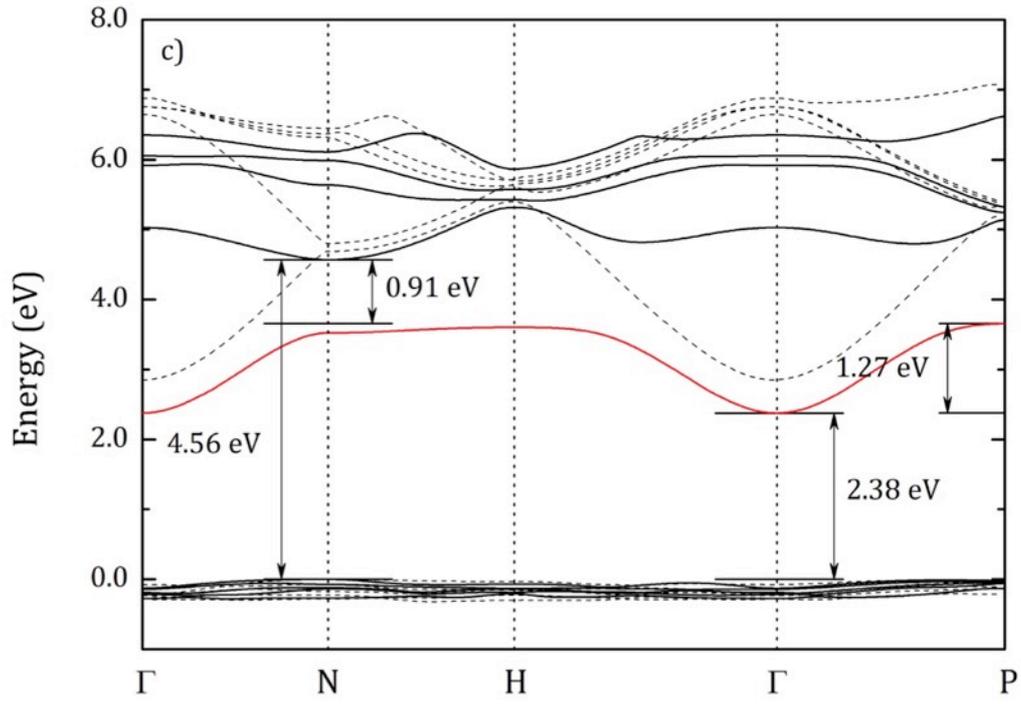

(c)

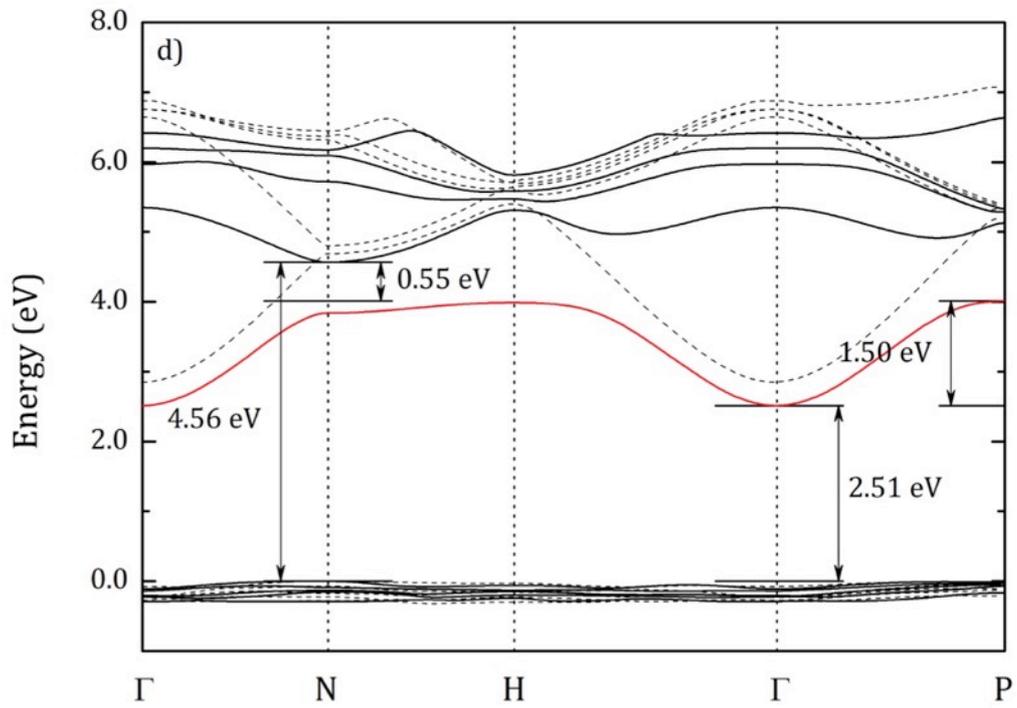

(d)

FIG. 6: Band structure of defective $In_2O_3$, along the same symmetry lines as in Fig. 5, for the stable configuration of defects: $H_i^{\cdot}$-AB (a), $H_O^{\cdot}$ (b), $v_O^{\cdot}$ (c) and $v_O^{\cdot\cdot}$ (d). The dashed lines correspond to the calculations without pSIC, and the arrows denote the bandgap in each case. The red lines correspond to the defect levels described in the text.

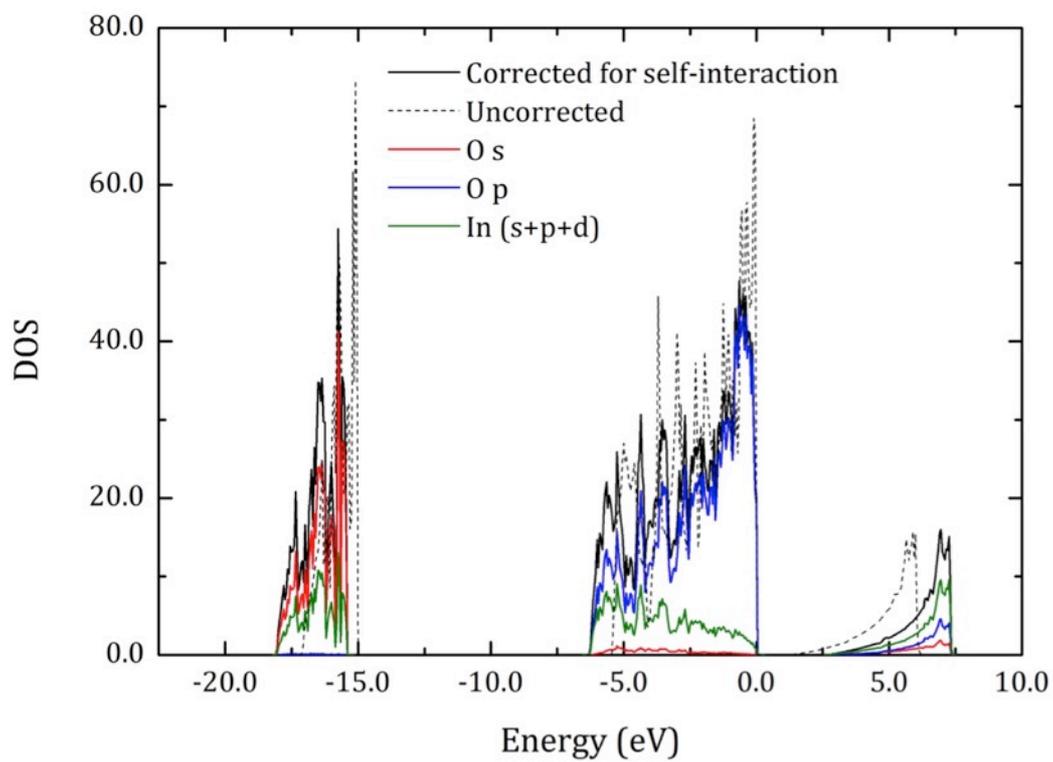

FIG. 7: Total (solid black lines) and orbital-projected DOS for defect-free $In_2O_3$. The individual projections on indium orbitals are not shown for clarity. The dashed line depicts the DOS calculated without pSIC.

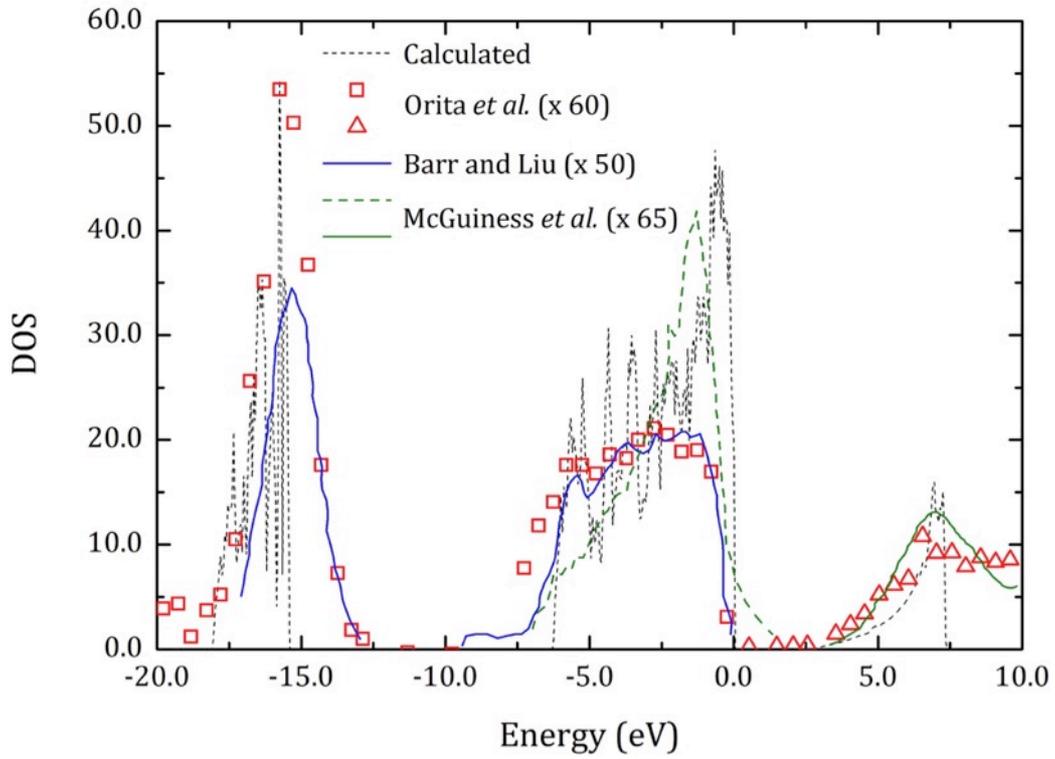

FIG. 8: Comparison of the calculated DOS for $In_2O_3$ (see Fig. 7) with experimental data published in the literature. The latter have been scaled for a proper comparison; the corresponding scale factors are shown in parentheses.

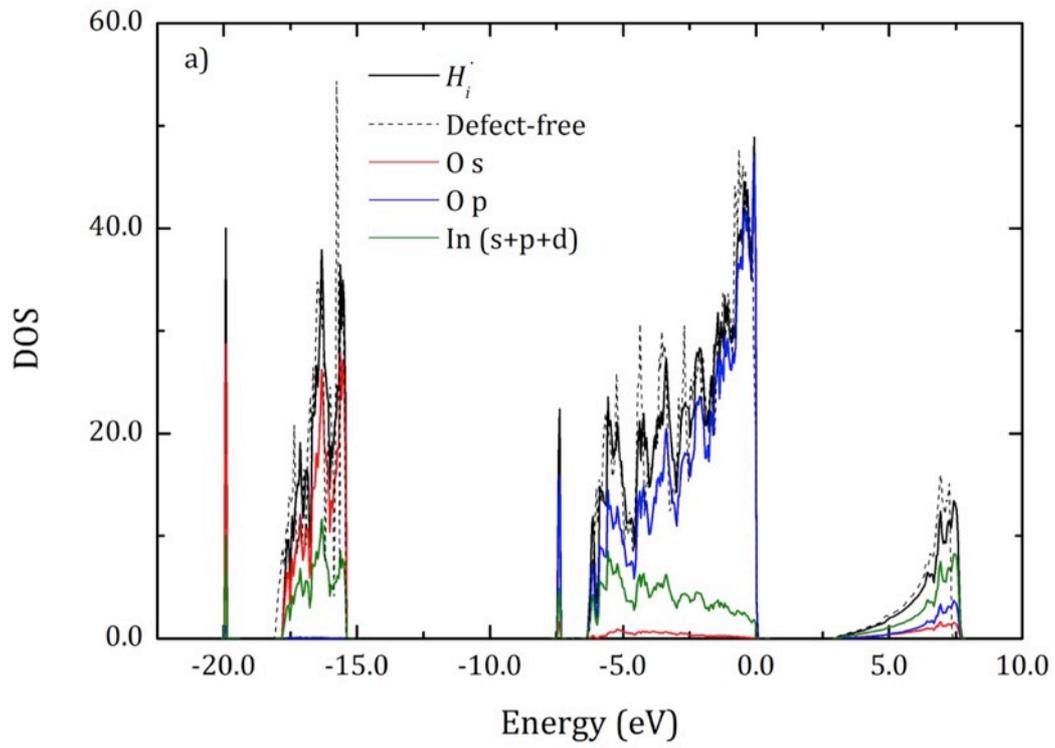

(a)

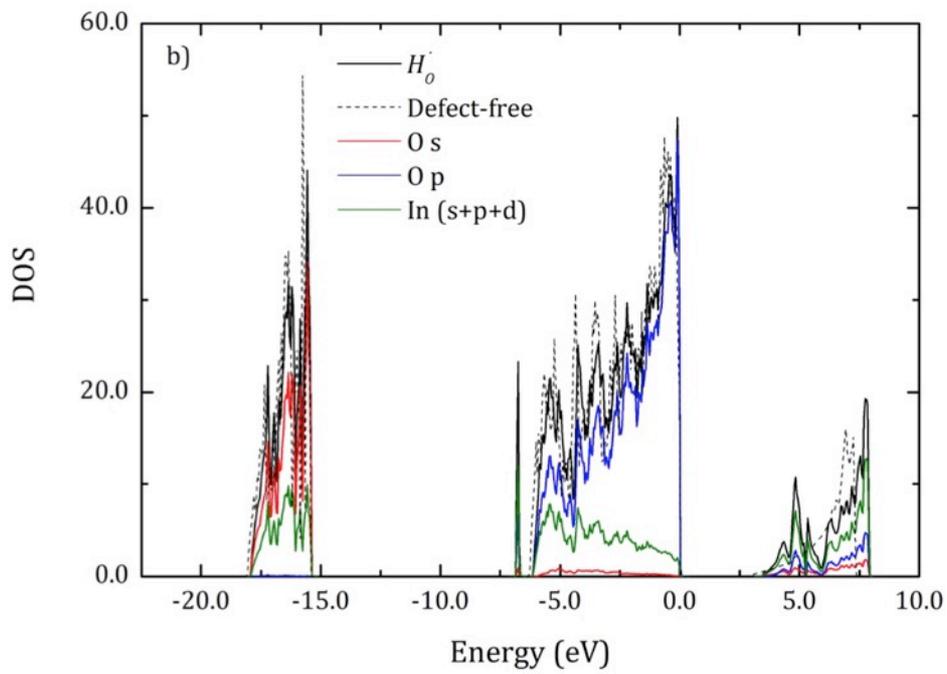

(b)

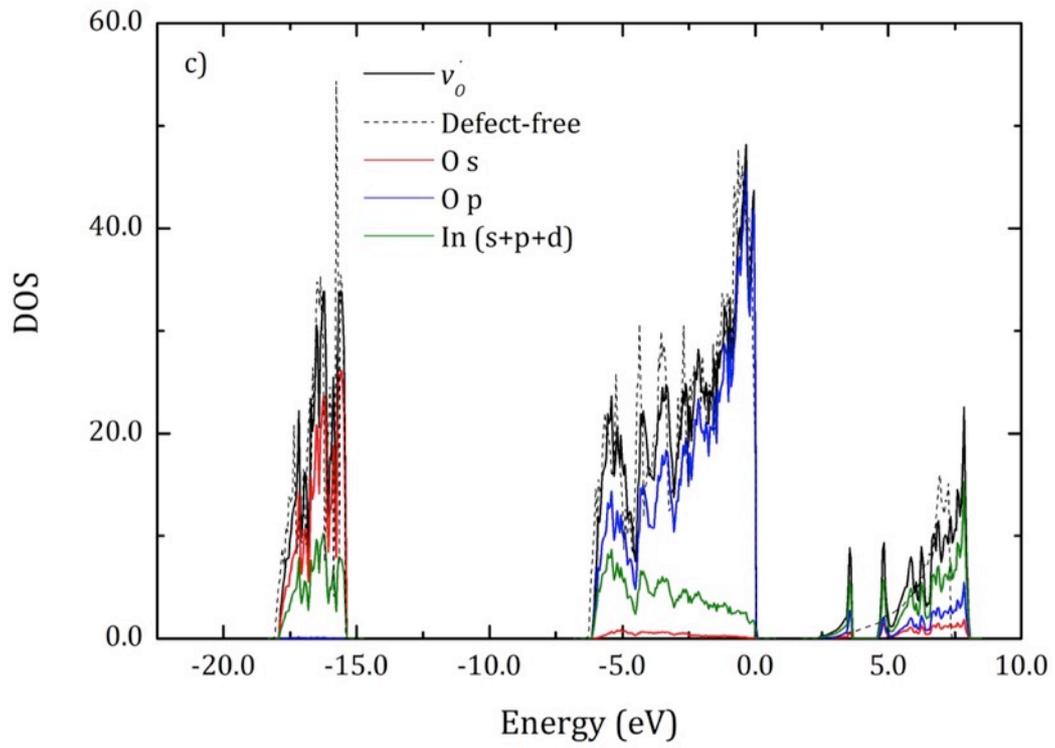

(c)

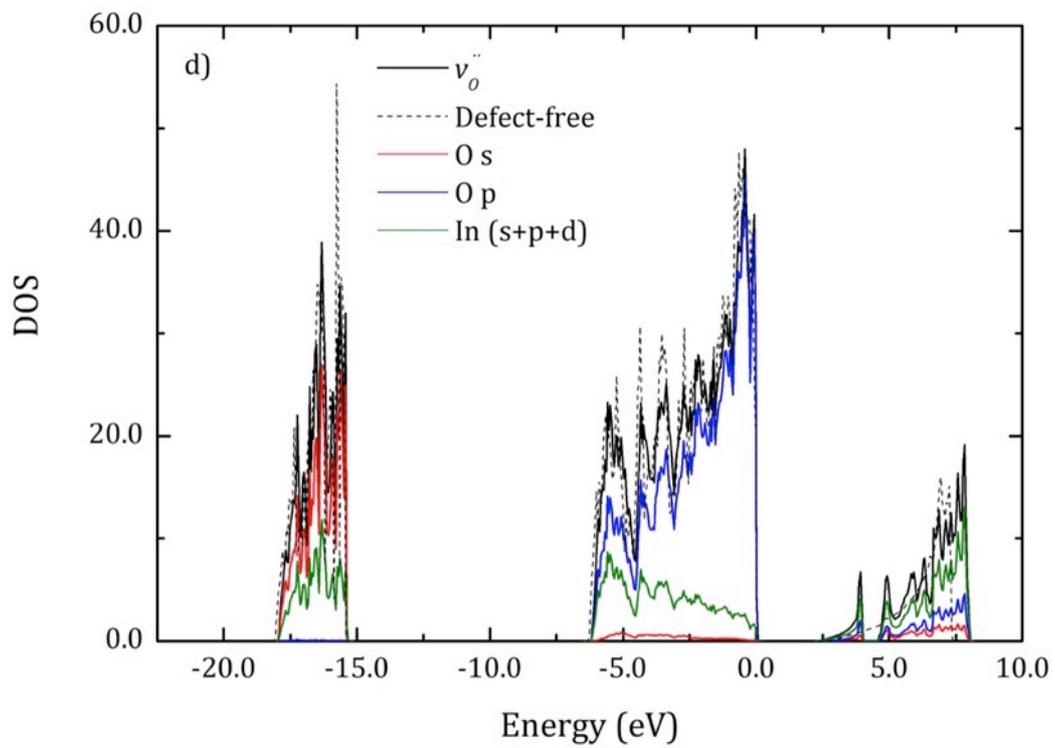

(d)

FIG. 9: Total and orbital-projected DOS for the stable configuration of defects in In$_2$O$_3$: $H_i^{\cdot}$ at AB (a), $H_O^{\cdot}$ (b), $v_O^{\cdot}$ (c) and $v_O^{\cdot\cdot}$ (d). The dashed lines depict the total DOS of the defect-free crystal.

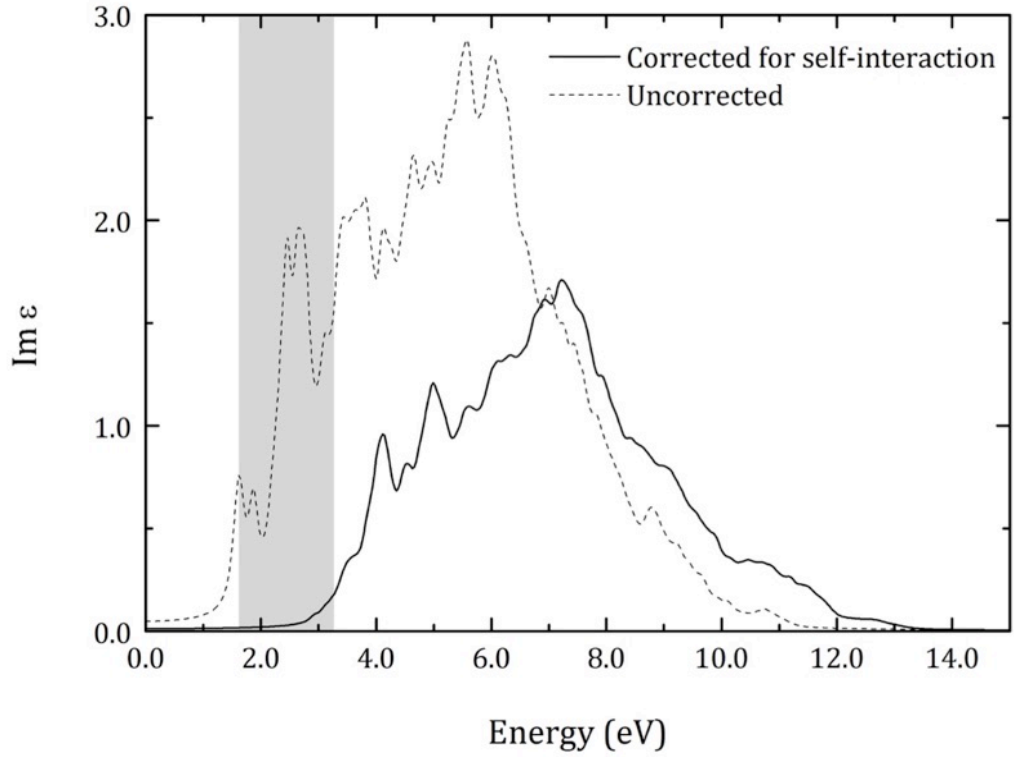

(a)

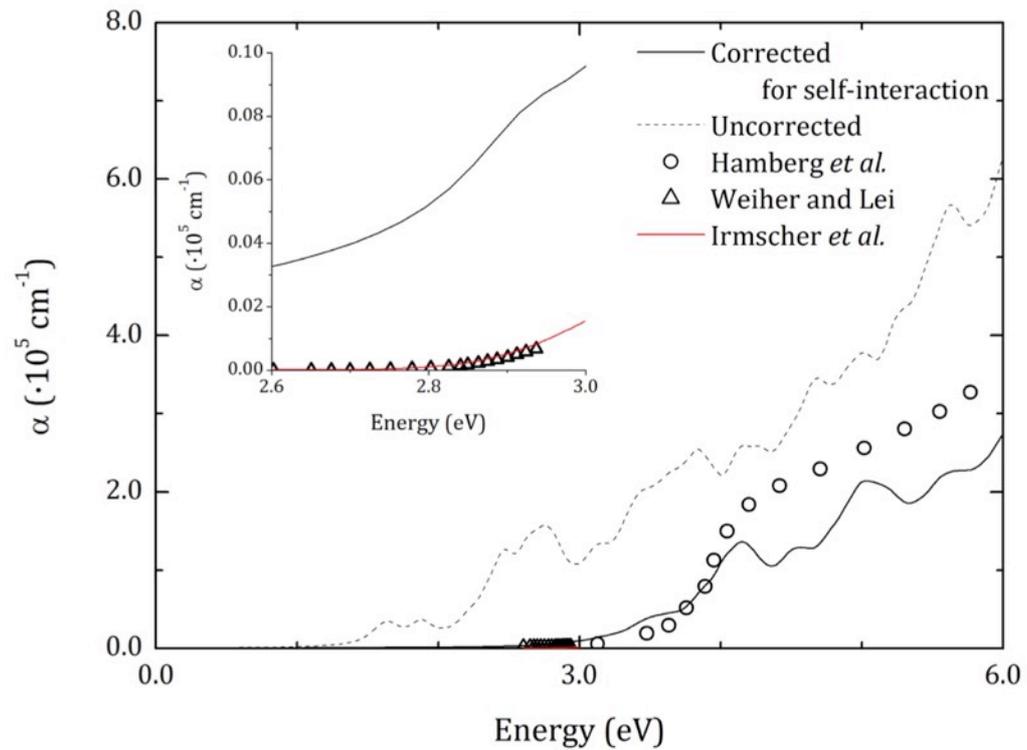

(b)

FIG. 10: a) Imaginary part of the dielectric function for defect-free In$_2$O$_3$ as a function of energy. The dashed line correspond to the calculation without pSIC, and the shadow region corresponds to the optical range; b) Absorption coefficient [Eq. (1)] as a function of energy at low energy, calculated with (solid line) and without (dashed line) pSIC, and comparison with experimental data available. The inset shows a detail of the plot within the range 2.6 – 3.0 eV.

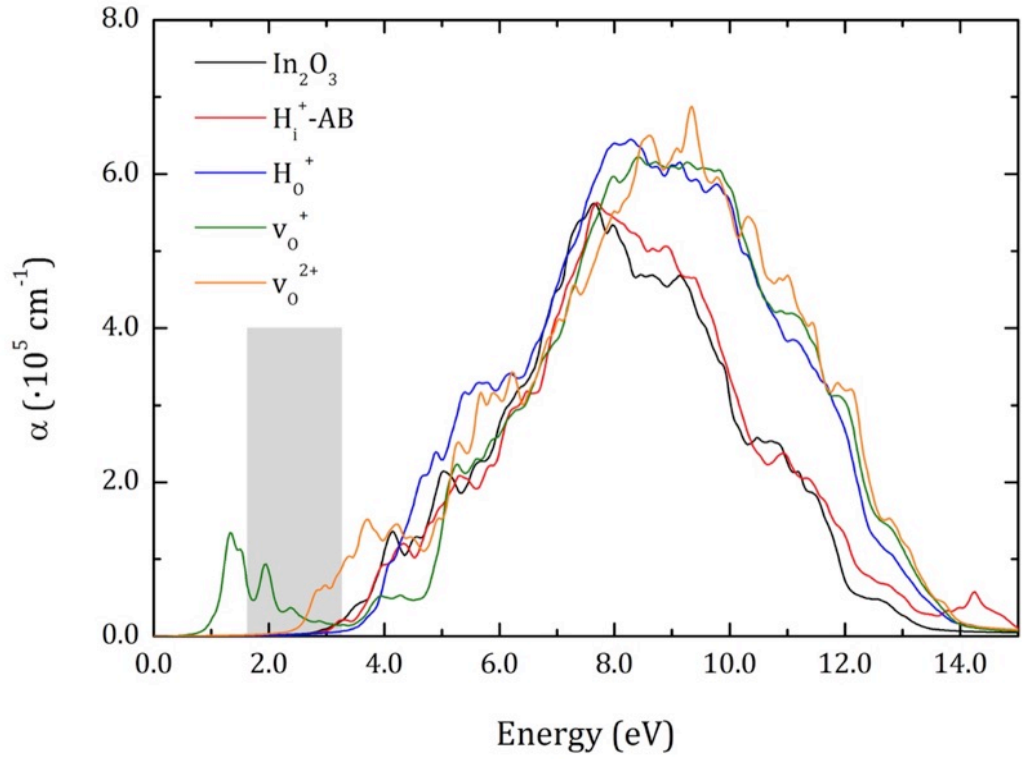

FIG. 11: Absorption coefficient [Eq. (1)] vs. energy for the stable configurations of defects in $In_2O_3$. The shadow region denotes the optical range.

**Table I**: Atomic coordinates in the supercell of defect-free $In_2O_3$.

| Atom | Wyckoff position | Coordinates | |
|---|---|---|---|
| | | Original | Relaxed |
| In-8$b$ | 8$b$ | (0.75, 0.25, 0.75) | (0.75, 0.25, 0.75) |
| In-24$d$ | 24$d$ | (0.9661, 0.5, 0.75) | (0.9625, 0.5, 0.75) |
| O | 48$e$ | (0.1103, 0.6545, 0.3824) | (0.1098, 0.6549, 0.3823) |